\documentclass[aps,prc,amsfonts,showpacs,floats,floatfix,twocolumn,nofootinbib,byrevtex]{revtex4}
\usepackage{ulem}
\usepackage{epsfig}
\usepackage{amsmath}
\usepackage{amsfonts}
\usepackage{color}
\usepackage{graphicx,slashbox}
\usepackage{multirow}

\begin{document}

\title{Exploration of a modified density dependence in the Skyrme functional}
\author{J. Erler$^1$, P. Kl\"upfel$^2$, P.--G. Reinhard$^1$}
\affiliation{
1) Institut f\"ur Theoretische Physik II, Universit\"at
  Erlangen-N\"urnberg, Staudtstrasse 7, D-91058 Erlangen, Germany
\\
2) Science Institute, University of Iceland,
   Dunhaga 3, IS-107 Reykjavik, Iceland}
\date{Received: date / Revised version: date}

\begin{abstract}
A variant of the basic Skyrme-Hartree-Fock (SHF) functional is
considered dealing with a new form of density dependence.  It employs
only integer powers and thus will allow
a more sound basis for projection schemes (particle number, 
angular momentum). 
We optimize the
new functional with exactly the same adjustment strategy as used in an
earlier study with a standard Skyrme functional. This allows direct
comparisons of the performance of the new functional
 relative to the standard one. We discuss various observables: 
bulk properties of finite nuclei, nuclear matter, giant resonances, 
super-heavy elements, and energy systematics. The new functional performs at
 least as well
as the standard one, but offers a wider range of applicability
(e.g. for projection) and more flexibility in the regime of high
densities.
\end{abstract}

\pacs{21.10.Dr, 21.10.Ft, 21.60.-n, 21.60.Jz}  
\maketitle

\section{Introduction}

Self-consistent mean-field models have grown to be a major tool in
the theoretical description of nuclei and nuclear dynamics.  There are
several approaches to nuclear SCMF from which the most widely used are
the Skyrme-Hartree-Fock (SHF) model \cite{Ben03aR}, the Gogny force
\cite{Dec80a}, and the relativistic mean-field model (RMF)
\cite{Rei89aR,Rin96aR}. All these models employ effective interactions
which provide reliable nuclear structure properties and low-energy
excitations at the level of a mean-field description. The general
strategy for constructing nuclear energy functionals for high-quality
calculations is to deduce the formal structure from principle
considerations \cite{Neg72a} and to rely on a phenomenological
adjustment of the model parameters.  There is a long history of SHF
development and optimization (for recent reviews see
\cite{Ben03aR,Sto07aR}) and yet, there is still demand for improvement
as can be seen from the many recent updates of the parameterizations,
for recent examples from SHF see e.g.
\cite{Sam02a,Ber05a,Ben07a,Klu09a,Sto09a}.  In spite of all these
activities, there remain several open problems which touch the
structure of present days SHF functionals \cite{Erl10a}. One such
point is the way in which the density dependence is modeled.  It used
to start from adding to the zero-range two-body term a ``three-body
term'' associated with a density dependence $\propto\rho^3$ where
$\rho$ is the local nucleon density \cite{Vau70a}. This was found to
be too restrictive and has been generalized to a profile
$\propto\rho^{2+\alpha}$ with $\alpha\sim$ 0.16--0.3 in order to
allow a better description of fission barriers and monopole
oscillations \cite{Bar82a}. Since then, this form for the one density
dependent term was carried forth unchanged. The reason is that most
empirical information on nuclear structure is associated with
densities near bulk equilibrium density $\rho_0\approx 0.16$
fm$^{-3}$. Systematic information in a wide range of
densities has not yet been acquired contrary, e.g., to the case of modeling liquid $^3$He
\cite{Ser91a,Wei91a}. Nonetheless, the traditional form
$\propto\rho^{2+\alpha}$ is an arbitrary guess, it is rather rigid
what extrapolation to the low density region (=weak binding) is
concerned, and in particular, the non-analytical behavior for
non-integer $\alpha$ turns out to be a great hindrance for
particle-number projection \cite{Dob07c,Erl10a}. It is the aim of this
paper to present and discuss an alternative form of the density
dependence for the SHF functional which is built in the form of
rational approximants using only integer powers of $\rho$. This will
then provide a SHF functional which can safely be used in projection
schemes, particularly for particle-number projection
\cite{Rin80aB}. Moreover, we will exploit the generalized form to
explore the effect of varied density dependence on bulk properties of
finite nuclei. The new SHF functional is calibrated and investigated
with the techniques of least-squares fitting and the basic data set as
developed in a previous large scale study \cite{Klu09a}. This allows a
direct comparison with the former results using the traditional
density dependence and it provides the same systematics of variations.

\section{The energy functional}

The SHF functional is formulated in terms of the following local
densities and currents (for details see \cite{Ben03aR,Klu09a} and
appendix \ref{app:dens}):
density $\rho_q$, kinetic-energy density $\tau_q$, spin-orbit density
$\mathbf{J}_q$, current $\mathbf{j}_q$, spin density $\sigma_q$, and
pair density $\xi_q$, where the index $q\in\{p,n\}$ labels protons and
neutrons; total densities are denoted as $\rho=\rho_p+\rho_n$ and
similarly for the other densities. Starting 
point is an expression of the total energy 
\begin{widetext}
\begin{subequations}
\begin{eqnarray}
  {E}
  & = & 
  {E}_\mathrm{kin}
            +\int d^3r\mathcal{E}_\mathrm{Sk}(\rho,\tau,\vec{j},\vec{J})
            +{E}_\mathrm{C}(\rho_p)
            +  E_{\rm pair}
            -{E}_\mathrm{cm}
\label{Etot} \\
   {E}_\mathrm{kin} &= &\int d^3r  \left(
      \frac{\hbar^2}{2m_p} \tau_p+  \frac{\hbar^2}{2m_n} \tau_n
   \right)
\label{Ekin} \\
   \mathcal{E}_\mathrm{Sk}
   &=&
   \mathcal{E}_\mathrm{Sk,tb}(\rho,\tau,\vec{j})
   +\mathcal{E}_\mathrm{Sk,ls}(\rho,\vec{J})
   +\mathcal{E}_\mathrm{Sk,dens}(\rho)
\label{ESk}\\
  \mathcal{E}_\mathrm{Sk,tb}
  &=& 
  \frac{b_0}{2}  \rho^2
  -\frac{b'_0}{2} \sum_q \rho_q^2
  +
  b_1 (\rho \tau - \vec{j}^2)
  - b'_1 \sum_q (\rho_q \tau_q - \vec{j}_q^2)
  -\frac{b_2}{2} \rho \Delta \rho 
  +\frac{b'_2}{2} \sum_q \rho_q \Delta \rho_q
\label{Esk-tb}  \\
  \mathcal{E}_\mathrm{Sk,dens}
  &=&
  \left[
     \frac{b_3}{2}  \rho^{2}
     - \frac{b'_3}{2}    \sum_q\rho_q^2
    \right]\rho^\alpha
  + \left[
     \frac{\tilde{b}_3}{2}  \rho^{2}
     - \frac{\tilde{b}'_3}{2}    \sum_q\rho_q^2
    \right]
     \frac{\alpha_\mathrm{inv}\rho+\beta_\mathrm{inv}\rho^2}
          {1+\alpha_\mathrm{inv}\rho}
\label{Esk-dens}\\
   \mathcal{E}_\mathrm{Sk,ls}
   &=&
   - b_4 \big( \rho\nabla\cdot\vec{{J}}
                 + (\nabla \times\vec{\sigma}) \cdot  \vec{j}
         \big)
   - b'_4 \sum_q \big( \rho_q(\nabla\cdot\vec{J}_q)
              + (\nabla \times\vec{\sigma}_q) \cdot  \vec{j}_q 
                  \big)
\nonumber
\\
   &&
     -
     \eta_\mathrm{tls}\Big(
       \frac{1}{16}(t_1x_1+t_2x_2)\vec{J}^2
       -
       \frac{1}{16}(t_1-t_2)\sum_q\vec{J}_q^2
     \Big)
\label{Eskyls}
\\
 {E}_C & = & \frac{1}{2}  e^2 \int d^3r \, d^3r' \rho_p(\vec{r})
              \frac{1}{|\vec{r}-\vec{r}'|} \rho_p(\vec{r}')
         -\eta_\mathrm{Cex}\frac{3}{4} e^2(\frac{3}{\pi})^\frac{1}{3} \int d^3r
                          [ \rho_p(\vec{r})]^\frac{4}{3}
              \quad,
\label{Ecoul}
\\
  E_{\rm pair}
  &=&
  \frac{1}{4} \sum_{q\in\{p,n\}}V_\mathrm{pair,q}
  \int d^3r \xi^2_q
  \big[1 -\frac{\rho}{\rho_{0,\mathrm{pair}}}\big]
  \quad,
\label{eq:ep2}
\\
      E_\mathrm{cm}
      &=&
\frac{1}{2mA} < \hat{P}^2_\mathrm{cm} >
      \quad,\quad \mbox{after variation.}
\label{eq:Ecm}
\end{eqnarray}
\end{subequations}
\end{widetext}
From this total energy it is straight forward to derive the
corresponding mean field equations by variation with respect to the
single-nucleon wavefunctions and the pairing amplitudes.

\subsection{The Skyrme energy functional}
The Skyrme functional in the narrower sense is given by the Skyrme
energy density $\mathcal{E}_\mathrm{Sk}$ comprised in the terms
(\ref{Esk-tb}--\ref{Eskyls}).  The parts (\ref{Esk-tb}) and
(\ref{Eskyls}) carry the zero-range two-body terms including effective
mass and spin-orbit couplings.  All density dependence is collected in
the term (\ref{Esk-dens}).  Its first contribution $\propto
{b}_3,{b}'_3$ represents the traditional density dependence.  The
second contribution adds the essential new piece where the density
dependence is modeled as a rational approximant.  Note that this term
employs only integer powers of $\rho$.  This new form has thus the
great advantage that it complies with particle-number projection.  The
idea is that this rational density dependence replaces the traditional
dependence $\rho^\alpha$. Thus we can switch the models by the
coefficients.  The traditional model is recovered by setting
$\tilde{b}_3,\tilde{b}'_3=0$.  The new model with rational density
dependence uses non-zero $\tilde{b}_3,\tilde{b}'_3$ and switches off
the traditional form with ${b}_3,{b}'_3=0$.  Both forms associate the
same density dependence to isoscalar and isovector terms.  One may
introduce more isovector freedom in the density dependent terms by
using all terms, i.e., $\tilde{b}_3,\tilde{b}'_3\neq 0$ together with
${b}_3,{b}'_3\neq 0$, but fixing $\alpha=1$ to maintain the crucial
feature that only integer powers of $\rho$ are used.  We will consider
this extended option to explore the impact of (isovector)
density dependence.

One may wonder why one needs such an involved rational form. A straight-forward 
approach would start with a mere Taylor expansion
for the density dependence, e.g. $\propto \rho^2 \left[
c_1\rho^1+c_2¸\rho^2+c_3\rho^3... \right]$. It turns out that this simple form
is  not able to reproduce the appropriate density dependence at low
densities. The form $\rho^\alpha$ with small $\alpha$ performs, in fact, very 
well in this regime which explains its great success in the development of 
the SHF functionals. It is clear that the non-analytical $\rho^\alpha$ resists 
a straightforward Taylor expansion. The rational approximant is much more 
flexible and allows a good fine tuning also in the low-density regime.

The spin-orbit part of the SHF functional (\ref{Eskyls}) contains the
tensor term with a switch factor $\eta_\mathrm{tls}$. The importance
of and the need for this term is still a matter of debate
\cite{Ben07a,Col07a,Klu09a}. The new parameterization developed here 
will employ the tensor terms, i.e. we deal with $\eta_\mathrm{tls}=1$\,.

The Skyrme energy as given above contains the minimum number of
time-odd terms which are needed to make the functional Galilean
invariant \cite{Eng75a}. There are many more time odd terms
conceivable at the level of given functionals
\cite{Pos85a,Dob95c,Ben03aR}. They are unimportant for even-even
nuclei and will be ignored here.

The parameters $b_i$, $b'_i$ used in the above definition of the
Skyrme functional are chosen to give a most convenient formulation of
the energy functional, the corresponding mean-field Hamiltonian and
residual interaction.  Traditional expressions for the Skyrme
functional use a different labeling. The relation to the standard
Skyrme parameters $t_i$, $x_i$
\cite{Eng75a,Bra85aR,Ben03aR,Sto07aR} is given in appendix
\ref{app:params}.

\subsection{Additional terms}

The direct Coulomb term employs the mere proton density $\rho_p$ and not the
charge density which would be obtained from folding with the nucleonic
charge distributions \cite{Fri82a}.  The difference is, in
principle, sizeable for heavy nuclei, but easily compensated for by
the other terms in the functional.  We keep the above form for reasons
of simplicity.

The Coulomb-exchange functional, the second term in (\ref{Ecoul}), has
been made switchable by the parameter $\eta_\mathrm{Cex}$.
Unfortunately, the LDA for Coulomb exchange invokes a non-integer
power of density which, again, inhibits particle-number
projection. Thus we will omit this contribution by using
$\eta_\mathrm{Cex}=0$ in the new parameterizations. Coulomb exchange
is a small contribution and the missing term will easily be taken up
by the other terms in the functional.

For the center-of-mass correction $E_\mathrm{cm}$, we use the
standard recipe as given in eq. (\ref{eq:Ecm}). Note the comment
``after variation'' therein. The $E_\mathrm{cm}$ is, in fact,
the expectation value of a two-body operator which makes the
mean-field equations in fully variational treatment rather
cumbersome. As $E_\mathrm{cm}$ is a small correction, we
evaluate it (non-variational) at the end of the 
mean-field calculation for then given single-particle wavefunctions.

The pairing functional (\ref{eq:ep2}) contains a continuous
switch, the parameter $\rho_{0,\mathrm{pair}}$ where volume pairing is
recovered for $\rho_{0,\mathrm{pair}}\longrightarrow\infty$ and
surface pairing for $\rho_{0,\mathrm{pair}}=0.16$ fm$^{-3}$.  We will
use $\rho_{0,\mathrm{pair}}$ as a free parameter and usually obtain a
form of the pairing functional which stays in between the extremes of
volume and surface pairing \cite{Klu09a}.

\subsection{Adjustment strategy and the variants in this survey}

The parameters of the new SHF functional are determined by
phenomenological adjustment as it is done for most SHF functionals.  We
employ exactly the same fit strategy and data as used in a recent large scale
survey \cite{Klu09a}. The data set has been selected to cover
spherical nuclei with negligible correlation effects \cite{Klu08a}
which amounts to isotopic and isotonic chains of semi-magic nuclei
(leaving out the mid-shell regions). As leading observables we
consider binding energy and key pattern of the nuclear charge
distribution: r.m.s. radius, diffraction radius and surface thickness
\cite{Fri86a}. For pairing properties we include the three-point gaps
as deduced from the odd-even staggering of binding energies and for
determining the spin-orbit parameters we take into account selected
spin-orbit splittings in doubly-magic nuclei. The actual adjustment
is performed with straightforward least-squares techniques. These
allow not only to determine the optimal parameters but also provide a
measure for the uncertainty on the parameters and deduced observables 
\cite{Klu09a,Naz10a}.

\begin{table}
\begin{center}
\begin{tabular}{|l|l|l|}
\hline
 {\em name} & {\em model} & {\em fit data}
\\
\hline
 SV-bas &  ${b}_3,{b}'_3\neq 0$; $\tilde{b}_3,\tilde{b}'_3=0$
        & 
       std, $a_{sym}$, $\kappa$,  $m^*/m$ 
\\
 RD-bas &  ${b}_3,{b}'_3=0$; $\tilde{b}_3,\tilde{b}'_3\neq 0$
        & 
       std, $a_{sym}$, $\kappa$,  $m^*/m$ 
\\
 RM-bas &  ${b}_3,{b}'_3\neq 0,\alpha=1$; $\tilde{b}_3,\tilde{b}'_3\neq 0$
        & 
       std, $a_{sym}$, $\kappa$,  $m^*/m$ 
\\
\hline
 SV-min &  ${b}_3,{b}'_3\neq 0$; $\tilde{b}_3,\tilde{b}'_3=0$
        & std
\\
 RD-min &  ${b}_3,{b}'_3=0$; $\tilde{b}_3,\tilde{b}'_3\neq 0$
        & std
\\
 RM-min &  ${b}_3,{b}'_3\neq 0,\alpha=1$; $\tilde{b}_3,\tilde{b}'_3\neq 0$
        & std
\\
\hline
\end{tabular}
\end{center}
\caption{\label{tab:names} Definition of the parameterizations.  The
  acronym RD stands for ``Rational Density dependence'' and RM for
  ``Rational and More density dependence''.  The entry ``std'' in the
  rightmost column indicates that the standard pool of fit data from
  finite nuclei is used \cite{Klu09a}. Further entries, if present,
  indicate the nuclear matter properties which had been added to the
  standard set: $a_{sym}=$ symmetry energy at bulk equilibrium (fixed
  at 30 MeV), $\kappa=$ isovector sum-rule enhancement factor (fixed
  at 0.4), $m^*/m=$ effective nucleon mass (fixed at 0.9).}
\end{table}
Table \ref{tab:names} shows the variants of the model studied in the
following. The parameterizations SV-bas and SV-min rely on the
standard SHF functional (note the $\tilde{b}_3,\tilde{b}'_3=0$) and
were presented in great detail in \cite{Klu09a}. SV-min results from a
straightforward fit to the data from finite nuclei. The ``min'' stands
for minimization of the quality measure $\chi^2$. This straightforward
fit turns out to leave large uncertainties on some nuclear matter
properties, and with it on nuclear giant resonance
excitations. Therefore a restricted fit was performed where some bulk
bulk properties (see last column of table \ref{tab:names}) were fixed
to commonly accepted values \cite{Ben03aR} which turn out to yield a
very good description of giant resonances in $^{208}$Pb. (The actual
values which were kept fixed can be read off from figure
\ref{fig:RD-nucmatprop}.) This led to SV-bas where ''bas'' stands for
base point because it was used in \cite{Klu09a} as the origin for a
systematic variation of bulk properties.

The new parameterizations take over the twofold strategy, free fit
``min'' and bulk restricted fit ``bas''. We consider first the new
density dependent term alone leading to the parameterizations RD-min
and RD-bas. A more flexible density dependence is explored by allowing
also the traditional density dependent term with fixed $\alpha=1$.
These are the parameterizations RM-min and RM-bas. Comparing the RD
forces with the RM forces allows to learn more about the impact of
isovector freedom in the density dependence.

We also compare with results from established SHF parameterizations:
SkM$^*$ as a widely used old standard \cite{Bar82a} which for the
first time managed to deliver acceptable surface energy and fission
barriers; SLy6 which had been developed with a bias to neutron rich
nuclei and neutron matter aiming at astrophysical applications
\cite{Cha98a}; SkI3 adding the freedom of an isovector spin-orbit
force to obtain an improved description of isotopic shifts of
r.m.s. radii in neutron rich Pb isotopes \cite{Rei95a}; BSk4 from
\cite{Gor03a} as representative of the series of fits to all available
nuclei \cite{Sam02a} having an effective mass of $m^*/m=0.92$ which is
best comparable to the new forces here.  Each one of these
parameterizations uses different sets of fit data, bias and
constraints. Nonetheless, they yield very similar results for nuclear
bulk properties, but differ in extrapolations and more detailed
observables \cite{Ben03aR}. Note that these traditional
parameterizations are taken as given. It is not possible to determine
their extrapolations errors in the results discussed later on.

The detailed parameters for the four new parameterizations are given
in table \ref{tab:params} in appendix \ref{app:params}. The extended
version RM-min and RM-bas have a negligible isoscalar parameter
$b_3$. A value of zero could be enforced without any significant change in quality
and other properties. However, the isovector parameter $b'_3$ is
large, accompanied by a similarly large change in the complementing
$\tilde{b}'_3$. The extended functional thus exploits its freedom, as
expected, to tune isovector properties.  The parameters
$\alpha_\mathrm{inv}$ and $\beta_\mathrm{inv}$ are very similar for
all four forces. The typical value of $\alpha_\mathrm{inv}\approx 15$
fm$^3$ means that the functional has a simple pole at
$\rho=-1/15\mathrm{fm}^{-3}\approx -0.067\mathrm{fm}^{-3}$. This means
that the functional is not strictly analytical. But this pole is
safely outside the range of possible single particle transition densities (whose absolute 
value is $\rho(\vec{r})$ but which can take any complex phase \cite{Dob07c,Rin80aB})
and thus the integrations in the projection schemes stay fully in the analytical regime.

\section{Results and discussion}

\subsection{Quality with respect to bulk observables}

\begin{figure}
\centerline{\epsfig{figure=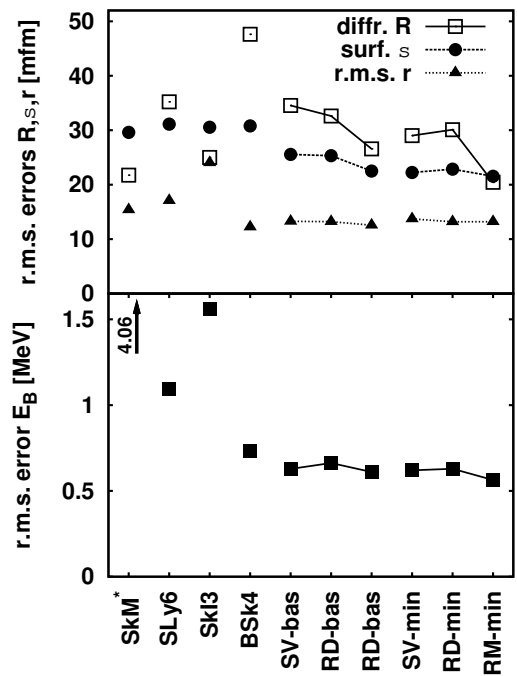,width=0.8\linewidth}}
\caption{\label{fig:RD-errors}
Average errors on nuclear bulk properties
for a variety of parameterizations as indicated.
The errors are obtained from averaging
over the reference set of SV-min \cite{Klu09a}.
Lower panel: errors on binding energy. Upper panel:
errors on  diffraction radius $R$, surface thickness
$\sigma$, r.m.s.radius $r$.
}
\end{figure}
Figure \ref{fig:RD-errors} shows the average errors for the four basic nuclear
bulk properties.  It is to be noted that the older parameterizations (SkM$^*$,
SLy6, SkI3, and BSk4) were tuned with different data and strategies than used
here.  This explains, e.g., the larger energy error for SkI3 which was fitted
to relative energy deviation and so has more bias to light nuclei, or the
larger error on the diffraction radius $R$ for BSk4 which was not included in
its fitting set. Nonetheless, comparing with the traditional
parameterizations, the progress in accuracy over the years 
becomes obvious. The parameterizations RD-bas and RD-min using the new density dependence provide a
description which is very much comparable to the SV-bas and SV-min which were
fitted to the same pool, of data but using the traditional density dependence
of the Skyrme functional.  The step to RM-min and RM-bas adds more flexibility
and necessarily should deliver an improved description. A reduction of errors
is indeed seen, but the gain is rather small.  From the perspective of quality
on bulk properties the new density dependence is neither harmful nor particularly
advantageous. It is an equivalent alternative having the formal advantage
to invoke only integer powers of density. It remains to been investigated how the new
parameterizations perform in other respects.

\subsection{Nuclear matter properties}

\begin{figure}
\centerline{\epsfig{figure=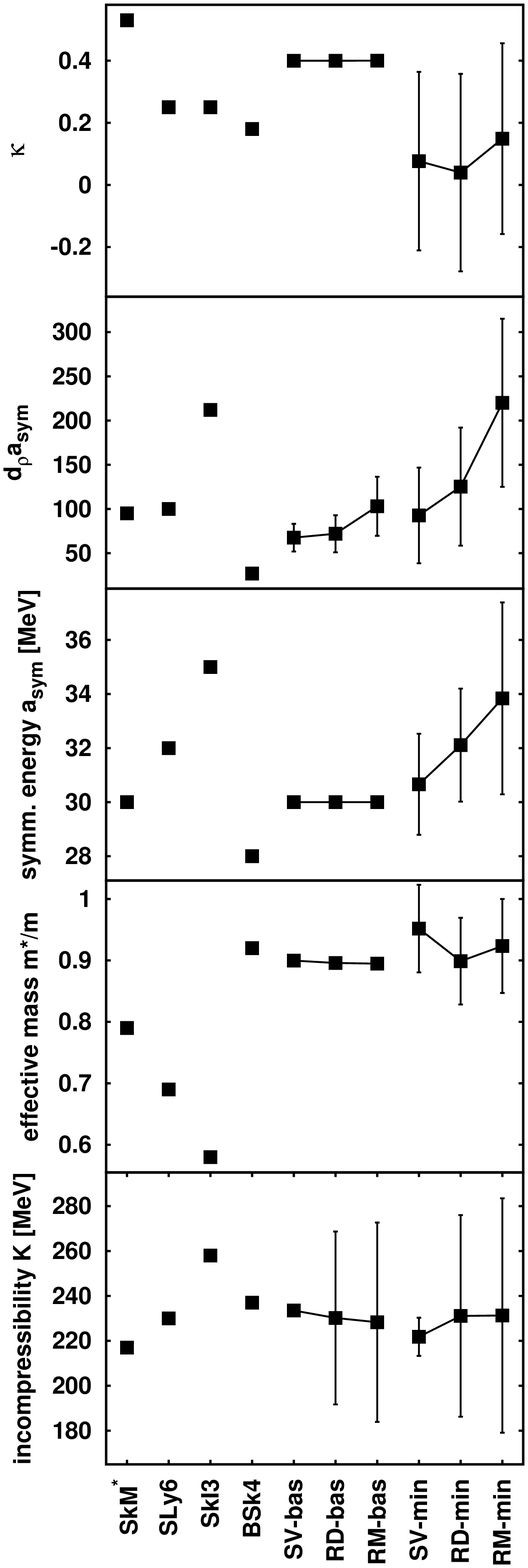,width=0.8\linewidth}}
\caption{\label{fig:RD-nucmatprop}
Nuclear matter properties (filled boxes) together
with their extrapolation uncertainty for the selected parameterizations.
}
\end{figure}
Figure \ref{fig:RD-nucmatprop} shows result for nuclear matter
properties.
The lowest panel shows the incompressibility
$K = 9\,\rho^2 \, \partial^2_\rho(E/A)$ where $(E/A)(\rho)$ is the bulk
binding of symmetric matter per nucleon.  The predictions are about
the same for all forces. This isoscalar property is rather well
determined by the wide span of nuclear sizes in all fit data.  The
extrapolation uncertainty is dramatically much larger for the new
density dependence (compare SV with RD and RM). This is due to the
larger flexibility of the new functional which has two parameters
instead of one in the old density model. The form of
density dependence can have a large influence on the bulk
incompressibility. The new functional would allow to adjust
much different values for $K$ if there is a need to do so.
The effective mass shows great variations for the traditional
parameterizations. The larger data sets of the new adjustments
(including BSk4) allow to determine its value in a rather narrow
window close to $m^*/m=0.9$. The ``bas'' parameterizations do not show
error bars because the value was fitted. The free fits in the ``min''
parameterizations confirm the choice with a rather small uncertainty.
This $m^*/m=0.9$ is fortunately also the value which allows a good
description of the giant quadrupole resonance.

All isovector properties (upper three panels) have large uncertainties
in the unconstrained fits (``min'' series). Moreover, their mean value
differs visibly from the constrained values which explains the
degrading of $\chi^2$ when switching to the constrained fits (see
figure \ref{fig:RD-errors}).  It is noteworthy that the extended
density dependence in RM-min leads also to some enhancement of
uncertainties in the isovector properties.  The effect, however, is
moderate in view of the freedom offered by the $b_3$, $b'_3$, terms in the
density-dependent contribution (\ref{Esk-dens}).

\begin{figure}
\centerline{\epsfig{figure=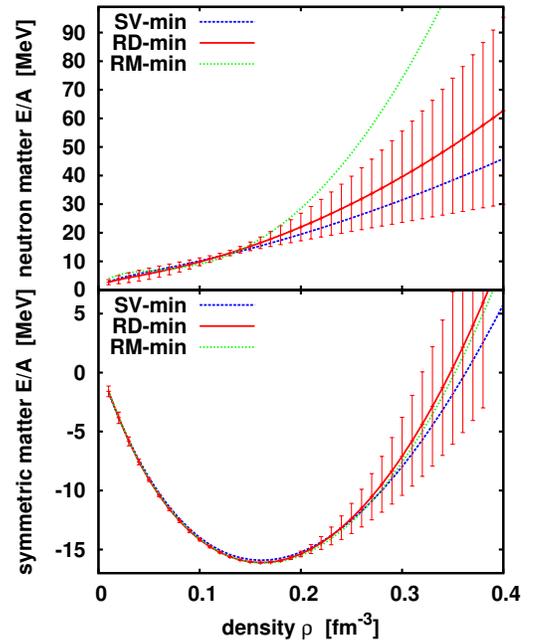,width=0.8\linewidth}}
\caption{\label{fig:RD-EoS}{\it(Color online)} Binding energy per nucleon
  in homogeneous nuclear matter as function of density
  Lower: for symmetric matter.
  Upper: for neutron matter.
  Shown are results for the ``min'' parameterizations as indicated.
  Those for the ``bas'' parameterizations are very similar.
  The values for RD-min are given together with errors bars
  representing the uncertainty in the extrapolation.
}
\end{figure}
Figure \ref{fig:RD-EoS} shows the binding energy curves in
homogeneous nuclear matter for SV-min using the traditional
density dependence and for RD-min as well as RM-min with the
new functional. All three parameterization predict practically 
the same energies for low and normal nuclear densities,
up to about $\rho=0.28$ fm$^{-3}$ for symmetric matter and
up to about  $\rho=0.15$ fm$^{-3}$ for neutron matter.
These are the densities which are explored in finite nuclei and it is
gratifying to see that the same information entering different models produce
the same predictions in the relevant range of density.
However, the extrapolation to high densities shows significant differences.
Very large effects are seen for RM-min in neutron matter due to the more
flexible isovector parameterization.
The difference in predictive power is nicely signaled by the
extrapolation uncertainties deduced from the least-squares techniques.
The error bands remain small in the region of low and normal nuclear
density and quickly enhance for higher densities.  The
figure shows only the uncertainties for RD-min. The error band has the
same trend for all parameterizations. It is in magnitude much the same
for RM-min, but a factor of five smaller for SV-min. This difference
indicates that the new density dependence, having two free parameters
instead of one, allows more flexibility to adjust the equation of
state in a broad density range. Unfortunately, the presently available
empirical data do not allow to exploit this flexibility. It may be
interesting to invoke results from ab-initio calculations to determine
the density dependence more uniquely.

\subsection{Giant resonances}

\begin{figure}
\centerline{\epsfig{figure=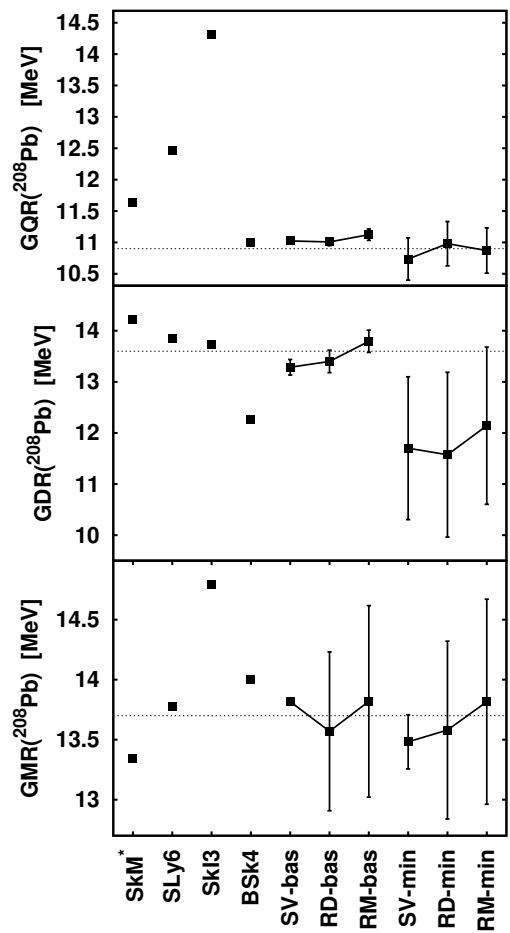,width=0.8\linewidth}}
\caption{\label{fig:RD-giantres}
Average peak energies of giant resonance excitations in
$^{208}$Pb. Lower panel: isoscalar giant monopole resonance (GMR);
middle panel: isovector giant dipole resonance (GDR), upper panel:
isoscalar giant quadrupole resonance (GQR). 
For the newly fitted forces, we show as error bars also the
uncertainties in the prediction.
The faint horizontal lines indicate the experimental resonance
positions (taken from \cite{Wou91aER,Vey70a,GDRdata}).
}
\end{figure}
Figure \ref{fig:RD-giantres} shows the peak frequencies for the three
basic giant resonances in $^{208}$Pb.  The resonance spectra were
computed with the random-phase approximation done self-consistently
with the same Skyrme interaction as was used for the ground state, for
technical details see \cite{Rei92a,Rei92b}. The peak positions were
deduced from an average over the resonance region covering $\pm 2$ MeV
around the peak.  The GMR comes out very similar and similarly correct
for all parameterizations.  New are the huge uncertainties for all new
parameterizations with the rational density dependence (RD and RM
forces). This results is much similar to the results for the
incompressibility in figure \ref{fig:RD-nucmatprop}. The similarity is
not surprising because it is known that incompressibility and GMR are
closely related to each other \cite{Bla80aR}.
The GDR behaves very similar with old and new density-dependence.
Unconstrained fits (rightmost three entries in the middle panel) yield
too low peak frequencies and large uncertainties while constrained
fits perform well by construction. There is an intimate relationship
between GDR and the symmetry energy $a_\mathrm{sym}$ as well as the sum
rule enhancement factor $\kappa$. The larger symmetry energies
together with lower $\kappa$ in figure \ref{fig:RD-nucmatprop}
correlate with the lower GDR energies in figure \ref{fig:RD-giantres}.
The GDR in $^{16}$O, however, remains troublesome as it was for all
previous SHF parameterizations \cite{Rei99a,Klu09a,Erl10a}.  The
resonance energy stays far below the experimental values. Even the
extended density dependence using $b_3,b'_3$ with their larger
isovector freedom does not resolve the puzzle.
The GQR (upper panel in figure \ref{fig:RD-giantres}) is known to have
clear relation to the effective mass $m^*/m$ \cite{Bra85aR}. This is
also found in the present results (when comparing with $m^*/m$ in
figure \ref{fig:RD-nucmatprop}).  The results for the RD and RM forces
are the same as for the SV forces.  This shows that there is no
influence from density dependence on the GQR.  The value $m^*/m\approx
0.9$ delivers a good reproduction of the GQR.

\subsection{Super-heavy elements (SHE)}

\begin{figure}
\centerline{\epsfig{figure=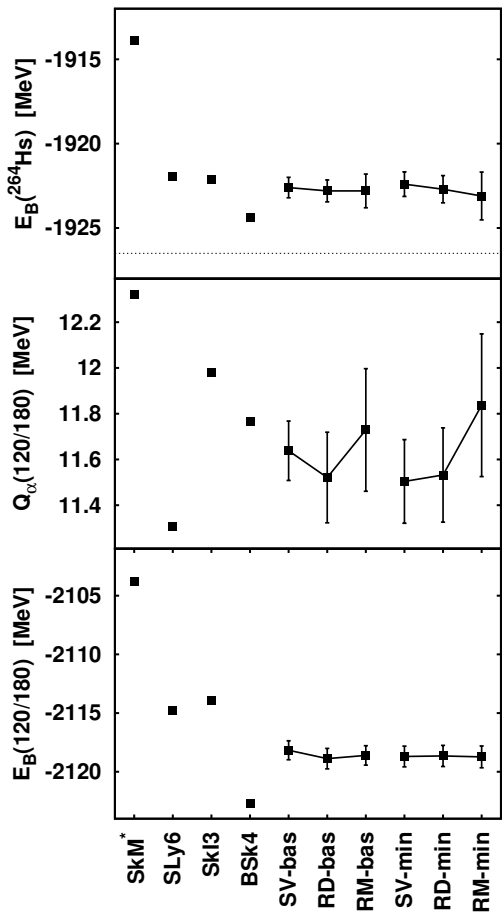,width=0.8\linewidth}}
\caption{\label{fig:RD-SHE} Properties of super-heavy elements (boxes)
  together with their extrapolation uncertainty for the selected
  parameterizations.  Lower: binding energy of the nucleus $Z=120$,
  $N=180$.  Middle: $Q_\alpha$ value for the nucleus $Z=120$, $N=180$.
  Upper: binding energy of $^{264}$Hs (including a correction for the
  rotational zero-point energy \cite{Klu09a}), the experimental value
  is indicated by a dashed horizontal line \cite{Hof00aER}.  }
\end{figure}
Figure \ref{fig:RD-SHE} shows results for super-heavy elements
(SHE). The results of the new parameterizations stay close to those of
SV-bas and SV-min. The new density dependence makes little difference,
mainly a slight growth of the uncertainty. Even then the
uncertainties remain small.  As a consequence, the mismatch of binding energy
in the known SHE $^{264}$Hs persists \cite{Erl10a}. We deduce from
the result that this mismatch is probably not related to an
insufficient density dependence and thus cannot be cured by more
flexible density profiles.

\begin{figure}
\centerline{\epsfig{figure=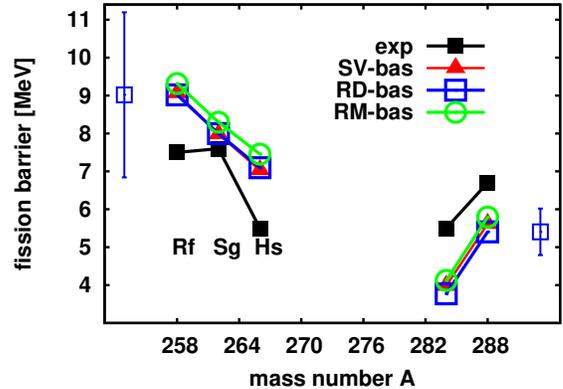,width=0.9\linewidth}}
\caption{\label{fig:barrier-RD-forces}{\it(Color online)} Fission barriers
  (filled boxes) of two groups of SHE for the selected
  parameterizations together with the experimental values.
  The extrapolation errors are very similar for all three forces.
  They are indicated for the lower group near the left margin and for
  the upper group near the right margin.
 }
\end{figure}
Another important observable in SHE are fission barriers. Figure
\ref{fig:barrier-RD-forces} shows results for barriers of two groups
of SHE, at the lower end Rf, Sg, Hs and more heavily Z=112,114 all
calculated with the new density dependences, with the traditional one,
and compared with experimental values \cite{Pet04,Itk02}.  The
predictions from the three forces shown in the figure are very much
the same. The parameterizations SV-min, RD-min, and RM-min would also
coincide and are not shown to keep the plot overseeable. The modified
density dependence does not make any difference on fission barrriers.
The agreement is generally satisfying.  Note, however, that the
experimentally observed trend of fission barriers cannot be reproduced
correctly by any SHF parameterization.  The barriers are somewhat
overestimated at the light side and underestimated in the heavier
group.  The same mismatch in the trend was found also for all
conventional forces like SkI3 or SLy6 and for fission life-times
\cite{Erl10a}.  The error bars indicated in the figure are as large as
the deviation, for the lighter SHE even larger. This nourishes some
hope that the trend could be resolved by appropriate retuning of the
forces.

\subsection{Energy systematics}

\begin{figure}
\centerline{\epsfig{figure=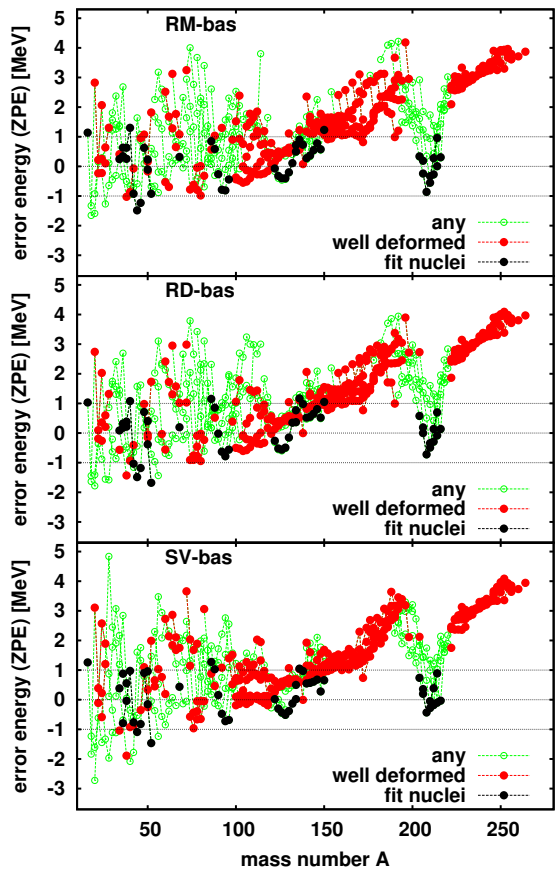,width=0.9\linewidth}}
\caption{\label{fig:RD-energies}
{\it(Color online)} Systematics of binding energies over all known even-even nuclei
for three parameterizations as indicated.
Fit nuclei are indicated by filled black circles, well deformed one by 
filled red circles, and all remaining one by open green circles.
}
\end{figure}
Figure \ref{fig:RD-energies} shows the systematics of errors of
binding energies $E_B^\mathrm{(SHF)}-E_B^\mathrm{(exp)}$ for all known
even-even nuclei and for the the forces SV-bas, RD-bas, and
RM-bas. The results for SV-min, RD-min, and RM-min look very similar
and are thus not shown here. The pattern in figure
\ref{fig:RD-energies} are practically the same for all three forces:
there is perfect agreement for the fit nuclei, there is good agreement
for all nuclei with $A<180$ when noting that the soft nuclei (the
remaining group, neither fit nuclei nor deformed) acquire additional
binding from correlation effects.  A systematic trend to under-binding
of heavy deformed nuclei is observed.  The mismatch already seen for
the case of $^{264}$Hs in figure \ref{fig:RD-SHE} was only one
spotlight out of this trend. Similar as in \cite{Klu09a} we have tried
to improve the performance for energies by including deformed nuclei
into the fit data set. This did improve the situation a bit, however,
at the price of a much degraded overall quality. We find again that
density dependence is not the key to the solution of this puzzle.

\section{Conclusions}

In this paper, we have proposed and investigated an alternative form for the
density dependent term in the Skyrme functional.  The traditional form
$\propto\rho^{2+\alpha}$ is replaced by a rational function $\propto
(\alpha_\mathrm{inv}\rho+\beta_\mathrm{inv}\rho^2)/
(1+\alpha_\mathrm{inv}\rho)$. The advantage of the new form is that it employs
only integer powers of $\rho$ and that it is analytical, except for one simple
pole at $\rho=-\alpha_\mathrm{inv}^{-1}$. The new form is thus applicable for
projection schemes as, e.g., angular momentum, center-of-mass projection and
particle number projection.We have also explored an extended form of the 
functional with one more density dependent term allowing for more 
flexibility in tuning isovector density-dependent terms. We have optimized 
the new functionals by adjustment to phenomenological data in standard manner 
and we have explored the properties of the newly developed functional in 
comparison to a couple of traditional Skyrme functionals. The straightforward 
methods of least-squares fitting allow also to deduce uncertainties on the 
predictions drawn from the fitted functionals.

The new forces provide at once an overall quality on standard nuclear bulk
properties (energy, r.m.s. radius, diffraction radius, surface thickness)
which is at least as good as the one obtained with the standard functional
using exactly the same fitting data. The extended parameterization even
provides a slight improvement. We have then checked the predictions for a
couple of other nuclear observables, nuclear matter at equilibrium, giant
resonance, super-heavy elements, fission, and energy systematics. The
predictions in all these observables are very similar for both new functionals
as compared to the recent fit within the standard Skyrme functional. 

Slight differences appear for the uncertainties in the extrapolations. The new
parameterizations, leaving more degrees of freedom, tend to produce larger
extrapolation errors in some observables. In particular, the errors on
incompressibility and giant monopole resonance are significantly larger.  The
extended new parameterization, having more isovector flexibility, produces
slightly larger uncertainties in isovector observables as, e.g., the symmetry
energy.

Large differences shows up for the equation-of-state of homogeneous
matter at large densities, particularly for neutron matter.  This is
not so surprising because the form of density dependence is
different. The complement of the observation is, in fact, very
satisfying. There is practically no difference seen in the range of
low and normal nuclear densities. This means that the adjustment to
finite nuclei has managed to accommodate the region of normal densities
in spite of the different form of the functionals.

\bigskip

Acknowledgment: 
This work was supported by BMBF under contract no. 06~ER~142D.

\appendix

\section{Densities and currents}
\label{app:dens}

The densities and currents are defined as
\begin{equation}
  \begin{array}{rcl}
   \rho_q(\vec{r})& = &\displaystyle
     \sum_{\alpha\in q} f_\alpha v_{\alpha}^2|\varphi_{\alpha}(\vec{r})|^2
  \\
   \tau_q(\vec{r})& = &\displaystyle
     \sum_{\alpha\in q} f_\alpha v_{\alpha}^2|\nabla\varphi_{\alpha}(\vec{r})|^2
  \\
   \vec{j}_q(\vec{r})& = &\displaystyle
   \Im{m}\left\{\sum_{\alpha\in q} f_\alpha v_{\alpha}^2 \varphi^+_{\alpha}(\vec{r})
              \nabla\varphi^{\mbox{}}_{\alpha}(\vec{r})\right\}
  \\
   \vec{\sigma}_q(\vec{r})& = &\displaystyle
     \sum_{\alpha\in q} f_\alpha v_{\alpha}^2
       \varphi^+_{\alpha}(\vec{r}) \hat{\vec{\sigma}}\varphi^{\mbox{}}_{\alpha}(\vec{r})
  \\
   \vec{J}_q(\vec{r}) & = &\displaystyle
    -i\sum_{\alpha\in q} f_\alpha v_{\alpha}^2
             \varphi_{\alpha}^+(\vec{r})
                \nabla \times \hat{\vec{\sigma}} \varphi^{\mbox{}}_{\alpha}(\vec{r})
  \\
   \xi_q(\vec{r})& = &\displaystyle
     \sum_{\alpha\in q}   f_\alpha  u_{\alpha} v_{\alpha}|\varphi_{\alpha}(\vec{r})|^2
  \end{array}
\label{rtj}
\end{equation}
where
$q$ labels the nucleon species with $q=p$ for protons and
$q=n$ for neutrons.
A density without an isospin label corresponds to a total density, i.e.
$\rho = \rho_p+\rho_n$ and similarly for the other densities.
The $v_\alpha$ and $u_\alpha$ are the standard BCS amplitudes,
$v_\alpha$ for occupation and $u_\alpha$ the complementing one for
non-occupation such that $v_\alpha^2+u_\alpha^2=1$. The $f_\alpha$ is
a phase-space weight which serves to provide a smooth cutoff of the
space of single-particle states included in pairing.
We use here a soft cutoff profile such as,
$
  f_\alpha
  =
  \left[1+
    \exp{\left((\varepsilon_\alpha-(\epsilon_F+\epsilon_{\rm cut}))
            /\Delta\epsilon\right)}
  \right]^{-1}
$
where typically $\epsilon_{\rm cut}=5\,{\rm MeV}$ and
$\Delta\epsilon=\epsilon_{\rm cut}/10$ \cite{Bon85a,Kri90a}.  This
works very well for all stable and moderately exotic nuclei.  For
better extrapolation ability away from the valley of stability, the fixed
margin $\epsilon_{\rm cut}$ may be modified to use a band of fixed
particle number $\propto N^{2/3}$ instead of a fixed energy band
\cite{Ben00c}.

\section{Details of the parameters}
\label{app:params}

\begin{table}
\begin{tabular}{|r|rrrr|}
\hline
            &  RD-bas    &   RD-min   &    RM-bas  &  RM-min \\
\hline
     $b_0$  & -1116.30   & -1116.34   & -1117.47   & -1116.13 \\
    $b'_0$      & -1127.16   & -977.786   & -1911.40   & -1705.29   \\
     $b_1$  &  15.0000   &  14.4966   &  15.0000   &  10.5256 \\
    $b'_1$      &  72.8588   & -18.9894   &  71.7330   &  16.5372   \\
     $b_2$  &  110.515   &  112.249   &  109.873   &  109.771 \\
    $b'_2$      & -284.614   & -317.996   & -81.9695   & -107.028   \\
    $b_3$  &            &            & -9.64712   &  9.95249 \\
    $b'_3$  &            &            & -4997.66   & -7137.87 \\
    $\alpha$  &            &            &  1.00000   &  1.00000 \\
     $\tilde{b}_3$  &  4364.93   &  4364.74   &  4395.28   &  4394.86 \\
$\tilde{b}'_3$  &  3325.54   &   4218.63  &  15058.7   &  16756.7   \\
$\alpha_\mathrm{inv}$&  15.1299   &  15.1269   &  15.2122   &  15.3502 \\
$\beta_\mathrm{inv}$ &  16.9485   &  17.0715   &  16.5759   &  17.1909 \\
    $b_4$   &  63.0960   &  55.6340   &  64.4335   &  61.3840 \\
    $b'_4$  &  36.3802   &  52.0100   &  33.5588   &  38.4206 \\
$V_\mathrm{pair,p}$  &  658.539   &  685.826   &  708.671   &  820.390 \\
$V_\mathrm{pair,n}$  &  601.751   &  654.283   &  642.556   &  740.647 \\
$\rho_{0,\mathrm{pair}}$  & 0.203601   & 0.194093   & 0.194622   & 0.176469 \\
\hline
$\frac{\hbar}{2m_p}$&\multicolumn{4}{|c|}{20.7498207} \\
$\frac{\hbar}{2m_n}$&\multicolumn{4}{|c|}{20.7212601} \\
$e^2$               &\multicolumn{4}{|c|}{1.44}\\
$\eta_{\mathrm{tls}}$   &\multicolumn{4}{|c|}{1}\\
$\eta_\mathrm{Cex}$ &\multicolumn{4}{|c|}{0}\\
\hline
\end{tabular}
\caption{\label{tab:params}
The parameters of the Skyrme functional (\ref{ESk}) for the
four new parameterizations.
}
\end{table}
Table \ref{tab:params} provides the parameters of the four new
parameterizations discussed in this paper. Dimension of length are given in
fm and dimensions of energy in MeV. This means:
$b_0,b'_0\leftrightarrow$ [MeV\,fm$^{3}$],
$b_1,b'_1,b_2,b'_2\leftrightarrow$ [MeV\,fm$^{5}$],
$b_3,b'_3,\tilde{b}_3,\tilde{b}'_3\leftrightarrow$ [MeV\,fm$^{6}$],
$b_4,b'_4\leftrightarrow$ [MeV\,fm$^{5}$],
$\alpha,\alpha_\mathrm{inv},\beta_\mathrm{inv}\leftrightarrow$ fm$^3$,
$e^2\leftrightarrow$ [MeV\,fm],
$\frac{\hbar^2}{2m}\leftrightarrow$ [MeV\,fm$^{2}$],
and
$\eta_{\mathrm{tls}}$ is a dimensionless switch factor.
 
The parameters $b_i$ and $b'_i$ used in the definition (\ref{ESk}) of
the Skyrme functional are chosen to give a most compact formulation of
the functional.  The traditional form of the Skyrme functional is
derived from an effective two-body interaction using force parameters
$t_i$ and exchange parameters $x_i$
\cite{Eng75a,Bra85aR,Ben03aR,Sto07aR}. There is a one-to-one
correspondence of the parameters which reads
\begin{equation}
  \begin{array}{rclcrc}
   b_0 & = &   t_0 (1+\frac{1}{2} x_0)
   \;\;, \\
   b'_0& = &   t_0 (\frac{1}{2}+x_0)       
   \;\;,   \\
   b_1 & = &  \frac{1}{4} \left[ t_1 (1+\frac{1}{2} x_1)+t_2 (1+\frac{1}{2} x_2)
              \right]
   \;\; , \\
   b'_1& = &  \frac{1}{4} \left[ t_1 (\frac{1}{2}+x_1)-t_2 (\frac{1}{2}+x_2)
              \right]                      
   \;\;,    \\
   b_2 & = &  \frac{1}{8} \left[ 3t_1 (1+\frac{1}{2} x_1)-t_2 (1+\frac{1}{2} x_2)
              \right]
   \;\; , \\
   b'_2& = &  \frac{1}{8} \left[ 3t_1 (\frac{1}{2}+x_1)+t_2 (\frac{1}{2}+x_2)
              \right]                      
   \;\;,    \\
   b_3 & = &  \frac{1}{4} t_3 (1+\frac{1}{2} x_3)
   \;\; , \\
   b'_3& = &  \frac{1}{4} t_3 (\frac{1}{2}+x_3)      
   \;\;,\\
   b_4 & = &  \frac{1}{2} t_4                
   \;\; . 
  \end{array}
\label{bdef}
\end{equation}
\bibliographystyle{apsrev}
\bibliography{biblio,reviews,add}

\end{document}